\documentclass[11pt]{article}
\usepackage{amsmath,amsthm,amssymb,amscd,amstext,amsfonts}

\usepackage{color,verbatim,graphicx,fullpage,url}
\usepackage{xcolor}
\usepackage{nicefrac}
\usepackage{fullpage,tikz,enumitem}
\usepackage{tikz-cd}
\usepackage{bm}
\usepackage{todonotes}
\usepackage{mathtools}
\usepackage{thm-restate}
\usepackage[numbers]{natbib}

\usepackage[breaklinks=true, linktocpage=true,]{hyperref}
\hypersetup{
    colorlinks = true,
    linkcolor={purple},
    urlcolor={purple},
    citecolor={blue!80!black}
}

\usepackage[nameinlink, noabbrev, capitalize]{cleveref}

\usepackage[utf8]{inputenc}
 
\Crefname{enumi}{Property}{Properties}

\theoremstyle{plain}

\newtheorem{thm}{Theorem}[section]
\newtheorem{theorem}[thm]{Theorem}

\crefname{atheorem}{Theorem}{Theorems}
\Crefname{atheorem}{Theorem}{Theorems}

\newtheorem{corollary}[thm]{Corollary}
\newtheorem{lemma}[thm]{Lemma}

\newtheorem*{claim*}{Claim}

\theoremstyle{remark}
\newtheorem{remark}[thm]{Remark}

\newcommand\norm[1]{\left|\!\left|#1\right|\!\right|}

\DeclareMathOperator{\E}{\mathbb{E}}  
\newcommand{\Edisp}{\operatorname{\mathbb{E}}\displaylimits}
\newcommand{\Prb}{\Pr}                                 
\DeclareMathOperator{\dist}{dist}                          
\DeclareMathOperator{\supp}{supp} 

\DeclareMathOperator{\RR}{R}

\newcommand{\eps}{\varepsilon}

\newcommand{\cU}{\mathcal{U}}

\newcommand{\authoraddress}[3]{%
  \par\medskip\noindent
  {\scshape #1}\par
  \noindent #2\par
  \noindent
  {\itshape E-mail address:}
  \href{mailto:#3}{\texttt{#3}}\par
}

\hypersetup{
    colorlinks = true,
    linkcolor = {purple},
    urlcolor = {purple},
    citecolor = {blue!80!black},
    pdftitle = {Communication Complexity of Disambiguations of Gap Hamming Distance},
    pdfauthor = {Yogesh Dahiya, Mika Göös, Nathaniel Harms, Hamed Hatami, Pooya Hatami, Shachar Lovett, Weiqiang Yuan},
    pdfsubject = {Communication complexity, randomized protocols, pseudodeterminism}
}

\title{Communication Complexity of Disambiguations of Gap Hamming Distance}

\author{
Yogesh Dahiya \quad
Mika G\"o\"os \quad
Nathaniel Harms \quad
Hamed Hatami
\\[0.5em]
Pooya Hatami \quad
Shachar Lovett \quad
Weiqiang Yuan
}
\date{}

\begin{document}
\maketitle
\vspace{-2.5em}
\begin{abstract}
We consider the Gap Hamming Distance problem in which Alice and Bob receive $n$-bit strings and must
decide whether their Hamming distance is at most $n/3$ or at least $2n/3$. This promise problem
has public-coin randomized communication complexity $O(1)$ since the players can estimate the
Hamming distance by comparing their strings on a few shared random coordinates. The required output
is undefined, or ``ambiguous'', on inputs with Hamming distance between $n/3$ and $2n/3$.

We prove that any disambiguation of this problem to a total function has randomized
communication complexity $\Omega(n^{2/3})$. As a corollary, we obtain an $O(1)$ versus
$\Omega(n^{2/3})$ separation between randomized and pseudo-deterministic communication complexity.
The proof was first found by an AI model and then reworked by the authors. 
\end{abstract}

\section{Introduction}
Denote the Hamming distance between two strings $x, y \in \{\pm1\}^n$ by
$\dist(x,y)$, and let $\gamma \in (0,1/2)$ be a fixed parameter. The
\emph{$\gamma$-Gap Hamming Distance problem} is the promise problem in which
Alice and Bob receive $x$ and $y$, respectively, and must distinguish between
the cases
\[
  \dist(x,y) \le \gamma n
  \qquad\text{and}\qquad
  \dist(x,y) \ge (1-\gamma)n.
\]
The randomized (public-coin) communication complexity
of this problem is $O_\gamma(1)$, since the players can
estimate $\dist(x,y)$ to sufficient accuracy by comparing their bits on a
constant-size set of shared random coordinates.

A \emph{disambiguation} of a promise problem is an extension of it to a total
Boolean function, that is, a total function that agrees with it on all inputs
that satisfy the promise. We prove that every disambiguation of
the $\gamma$-Gap Hamming Distance problem has large randomized
complexity.  
\begin{theorem}\label{thm:main}
Let $\gamma \in (0,1/2)$ be fixed. Every total function
$F \colon \{\pm1\}^n \times \{\pm1\}^n \to \{0,1\}$ satisfying
\[
  F(x,y) =
  \begin{cases}
    0 & \text{if } \dist(x,y) \le \gamma n, \\
    1 & \text{if } \dist(x,y) \ge (1-\gamma)n
  \end{cases}
\]
has randomized communication complexity $\Omega_\gamma(n^{2/3})$.
\end{theorem}

Previously, the best known lower bound for this problem was $\Omega(\log\log n)$,
due to~\cite{blondal2026borsuk}. A linear lower bound
$\Omega_\gamma(n)$ is conjectured in~\cite{Goos2026}.

\paragraph{Pseudo-deterministic communication.}
Our result is partly motivated by the study of pseudo-determinism in communication complexity.
When solving a search problem, a randomized algorithm that is only required to be correct with
high probability may nevertheless output different (but still correct) outputs between different
independent executions. This may be undesirable behaviour, so we may demand a randomized algorithm
whose outputs are consistent (with high probability) between different executions. Such an algorithm
is called \emph{pseudo-deterministic}. Pseudo-determinism was introduced by Gat and
Goldwasser~\cite{GatG11} (but also independently by Huynh and Nordstr\"om~\cite{HN12} in the context of proof complexity) and has since been studied across a wide range of computational models,
including learning algorithms~\cite{OS18,ILPS22}, sequential and parallel algorithms~\cite{GGR13,GG17}, average-case and approximation
algorithms~\cite{Hol17,DPV18}, interactive proofs~\cite{GGH18}, low-space algorithms~\cite{GL19}, and streaming algorithms~\cite{MR4048182}. 
The central question in this line of research is to what extent pseudo-determinism can be separated from
general randomized computation.

Formally, a communication \emph{search problem} is a relation
$\mathcal{R} \subseteq \{\pm1\}^n \times \{\pm1\}^n \times \mathcal{Z}$: Alice
and Bob receive $x$ and $y$, respectively, and must output some $z$ with
$(x,y,z) \in \mathcal{R}$. A \emph{pseudo-deterministic} protocol $\Pi$ for $\mathcal R$ is a
randomized protocol such that, for every
$(x,y)$, there is a fixed value $h(x,y)$ with $(x,y,h(x,y)) \in \mathcal R$ such that $\Pr[\Pi(x,y)
= h(x,y)] \geq \tfrac23$. We call
$h(x,y)$ the \emph{canonical output}.

Although we aim to compare pseudodeterministic protocols to randomized protocols, there is no
standard definition of what it means for a randomized protocol to solve a search problem. For \emph{functions}, the standard definition is that the algorithm is required to output the
unique correct answer with probability at least $2/3$, and this choice of constant $2/3$ is
essentially arbitrary because we can efficiently amplify the success probability to any other
constant. We cannot always amplify the success probability for search problems (because the
algorithm may output different correct answers on different executions), so we cannot simply choose
a fixed constant probability of success. The paper~\cite{Goos2026} discusses these nuances in more detail and
surveys various definitions in the literature.  However, for this paper, we can say
that a randomized protocol $\Pi$ solves $\mathcal{R}$ if~$\Pr[(x,y,\Pi(x,y)) \in \mathcal{R}] \ge
\frac23$ for every $(x,y)$. This is justified by the fact that the problems we study do allow
efficient amplification of the success probability.

The problem of separating randomized and pseudo-deterministic communication
complexity was posed in~\cite{MR4287109}. The first separations, $O(1)$
versus $\Omega(\log\log n)$~\cite{blondal2026borsuk} and $O(\log n)$ versus
$\Omega(\sqrt{n})$~\cite{Goos2026,Wat2026}, were obtained only recently.  \cref{thm:main} improves on
both of these results, giving a separation of $O(1)$ versus $\Omega(n^{2/3})$. This is because the
$\gamma$-Gap Hamming Distance problem is a relation (where the relation $\mathcal R$ allows either 0
or 1 outputs on distances between $\gamma n$ and $(1-\gamma)n$), and a pseudodeterministic algorithm
for this problem is equivalent to a randomized algorithm that computes a disambiguation.

Furthermore, as an application of our main result, we obtain an
$O(1)$ versus $\Omega(n^{2/3})$ separation for Hamming-distance
estimation: for a fixed $\varepsilon \in (0,1/2)$, the players must output a
number $t$ satisfying
\begin{equation}\label{eq:HammingApprox}
  |\dist(x,y) - t| \le \varepsilon n.
\end{equation}
Estimating the Hamming distance from $O_\varepsilon(1)$ shared random
coordinates solves this problem with~$O_\varepsilon(1)$ communication. Now
suppose that $\Pi$ is a pseudo-deterministic protocol with canonical output
$h(x,y)$, and consider the total function
\[
  F(x,y) \coloneqq \mathbf{1}[h(x,y) \ge n/2].
\]
Note that $F$ satisfies the
conditions of \Cref{thm:main} with $\gamma = \frac{1/2 - \varepsilon}{2}$.
Moreover, running $\Pi$ and outputting $1$ if and only if its output is at
least $n/2$ computes $F$ with error at most $1/3$. We therefore obtain the
following.

\begin{corollary}\label{cor:separation}
For every fixed $\varepsilon \in (0,1/2)$, the search problem
\eqref{eq:HammingApprox} has randomized communication complexity
$O_\varepsilon(1)$ and pseudo-deterministic communication complexity
$\Omega_\varepsilon(n^{2/3})$.
\end{corollary}

\paragraph{Parity decision trees.}
For $z\in\{\pm1\}^n$, let $|z|\coloneqq|\{i:z_i=-1\}|$. For
$\gamma\in(0,1/2)$, the $\gamma$-Gap Majority problem is the promise
problem $f\colon\{\pm1\}^n\to\{0,1,*\}$ defined by
\begin{equation}\label{eq:gap-majority}
  f(z)=
  \begin{cases}
    0 & \text{if } |z|\le\gamma n,\\
    1 & \text{if } |z|\ge(1-\gamma)n,\\
    * & \text{otherwise}.
  \end{cases}
\end{equation}
The $\gamma$-Gap Hamming Distance problem is its {\sc xor}-lift
$F(x,y)=f(x\odot y)$, where $\odot$ denotes coordinatewise multiplication,
the $\{\pm1\}$ representation of {\sc xor}.

Each parity query to $x\odot y$ can be simulated with two bits of
communication by exchanging the parities of $x$ and $y$ on the queried
set. Thus a randomized parity decision tree for any disambiguation of
$\gamma$-Gap Majority yields a randomized communication protocol for a
disambiguation of $\gamma$-Gap Hamming Distance, with at most twice the
cost. By \Cref{thm:main}, we obtain the following.

\begin{corollary}\label{cor:parity}
For every fixed $\gamma\in(0,1/2)$, every disambiguation of the
$\gamma$-Gap Majority problem has randomized parity decision tree
complexity $\Omega_\gamma(n^{2/3})$.
\end{corollary}

This improves Gavinsky's lower bound of $\Omega_\gamma(\sqrt{n})$ for
the same problem~\cite{Gavinsky}. In fact, our proof of \Cref{thm:main} follows
his iterative strategy.

\paragraph{AI methodology.}
An AI model first produced a proof of an $\Omega(\sqrt{n})$ lower bound, and
after further generic prompts, it improved the bound to $\Omega(n^{2/3})$.
The authors extracted the key ideas from these AI-generated proofs and used
them to write an argument in the form they found most
natural and intuitive.
The exposition in this note was drafted by the authors, with AI tools used
for editing, typesetting, and spell-checking.

\paragraph{Notation.}
We write $\RR_\eps(F)$ for the public-coin randomized communication complexity of $F$ with error at most $\eps$ on every
input. We write $\cU_m$ for the uniform distribution on $\{\pm1\}^m$,
and $\operatorname{Ber}(\beta)$ for the distribution on $\{\pm1\}$ that
assigns probability $\beta$ to $1$ and $1-\beta$ to $-1$.

We use $H$ to denote the usual Shannon entropy. The \emph{min-entropy} of a distribution $\mu$ on
a finite set $\Omega$ and its \emph{deficit} are defined respectively as (all logarithms are to base~$2$) 
\[H_\infty(\mu) \coloneqq -\log \norm{\mu}_\infty \quad \text{ and }\quad   D_\infty(\mu) \coloneqq \log (|\Omega|  \norm{\mu}_\infty).\] 
For $\rho\in[-1,1]$, a random variable $(X,Y)$ taking values in
$\{\pm1\}^m\times\{\pm1\}^m$ is \emph{$\rho$-correlated} if each
marginal is uniform, the pairs $(X_i,Y_i)$ are independent, and
$\E[X_iY_i]=\rho$ for every $i$. For $x\in\{\pm1\}^m$, we write
$N_\rho(x)$ for the conditional distribution of $Y$ given $X=x$: independently in
each coordinate, $Y_i=x_i$ with probability $(1+\rho)/2$ and $Y_i=-x_i$
with probability $(1-\rho)/2$.

\section{Proof overview}
\label{sec:main_argument}
We prove \Cref{thm:main} by contradiction. Suppose that $F$ has a randomized
protocol with small communication cost. Replacing $F(x,y)$ by
$1-F(x,-y)$ if necessary, we may assume that
\[
  \Edisp_{\cU_n\times\cU_n}F\ge\frac12.
\]
We follow the iterative strategy of Gavinsky~\cite{Gavinsky}. By exploiting the low-cost protocol
for $F$, we progressively grow a set $K$ of coordinates and force Alice's and Bob's inputs to
agree on most of the coordinates in $K$, while maintaining $\E F \ge 1/2$. Eventually, every
pair of inputs consistent with the forced coordinates $K$ has Hamming distance $< \gamma n$, so $F$
is guaranteed to be $0$, which contradicts the maintained property $\E F \ge 1/2$.

More precisely, we maintain a product distribution $\mu \times \nu$ on the
inputs, a set $K$ of processed coordinates, and a reference string
$a \in \{\pm1\}^K$, with two properties:
\begin{enumerate}[label=(\roman*)]
  \item $\E_{\mu \times \nu} F \ge 1/2$;
  \item for every $(x,y)$ in the support of $\mu \times \nu$, both 
    $\dist(x_K, a)$ and $\dist(y_K, a)$ are small (and therefore $\dist(x_K, y_K)$ is
    small).
\end{enumerate}
We start with uniform inputs and $K$ empty. Each round adds
new coordinates to~$K$, in two steps:
\begin{enumerate}
  \item \emph{Planting a mostly agreeing block (\Cref{lem:planting}).} We choose a
    large block $I$ outside $K$ and a reference string $w$ on it, and restrict
    both inputs to be close to $w$ on $I$, while keeping $\E F > 1/4$. We
    then add $I$ to $K$ and extend $a$ by $w$.
  \item \emph{Restoring the expectation (\Cref{lem:restoring-expectation}).}
    We use the low-cost protocol for $F$ to find a rectangle (leaf of the protocol) of sufficiently
    large probability on which $\E F \ge 1/2$. Conditioning on this rectangle
    restores the expectation.
\end{enumerate}
Eventually, every pair of inputs in the support differs in at most
$\gamma n$ coordinates. Then $F = 0$ on the entire support, contradicting~(i).

To see the idea,
suppose first that $X$ and $Y$ are uniform. We replace them
with a $\rho$-correlated pair, obtained by choosing a random
set $I\subseteq[n]$, including each coordinate independently
with probability $\rho$, and sampling $X,Y$ uniformly subject
to $X_I=Y_I$. The following theorem shows that, for
$\rho=\Theta(d^{-1/2})$, this  changes the acceptance
probability of the protocol by only a small constant.

\begin{theorem}[Girish--Sinha--Tal--Wu, {\cite[Theorem~1.5]{GSTW}}]
\label{thm:gstw}
For every randomized protocol $\Pi$ with output in $\{0,1\}$ that
communicates at most $d$ bits, and every $\rho\in[-1,1]$,
\[
  \Bigl|\Prb_{X,Y\sim\cU_m}[\Pi(X,Y)=1]
  -\Prb_{X\sim\cU_m,\,Y\sim N_\rho(X)}[\Pi(X,Y)=1]\Bigr|
  \le O\bigl(|\rho|\sqrt d\,\bigr).
\]
\end{theorem}
Since the protocol has small error, the expectation of $F$ also changes by
only a small constant. By averaging, we can then fix a block
$I$ of size $\Omega(\rho n)$ and a common value
$w=X_I=Y_I$ while maintaining $\E F>1/4$.

After we restrict to a rectangle to restore the expectation,
the inputs are no longer uniform, so \Cref{thm:gstw} does
not apply directly. To handle this, we use a coupling
(\Cref{lem:coupling}): any random string $X$ of high
min-entropy can be coupled with a uniform random string
$U$ so that $X$ and $U$ are close in expected Hamming
distance. Applying this separately to Alice's and Bob's
inputs gives uniform proxy strings $U$ and $V$, and we can apply the correlation argument
above to $(U,V)$ instead of inputs. Since each
input is close to its proxy string, both inputs are close to $w = U_I = V_I$
on the block $I$.

The coupling also guarantees that, after fixing $(I,w)$
and discarding a small exceptional set, the inputs on the
remaining coordinates still have high min-entropy.
Moreover, conditioned on $(I,w)$, Alice's and Bob's inputs
are independent.

\section{Coupling with the uniform distribution}
 
The following lemma couples a distribution $\mu$ on $\{\pm1\}^m$ of high
min-entropy with the uniform distribution on $\{\pm1\}^m$, using a sequential
maximal coupling due to Marton~\cite{Marton}. The
resulting pair $(X,U)$, with $X \sim \mu$ and $U$ uniform, has two
properties: (a)~$X$ and $U$ are close in expected Hamming distance, and
(b)~for every set $I$ of coordinates, revealing $U_I$ does not significantly
increase the probabilities of individual values of $X_{\bar I}$, except on a
small exceptional set.

\begin{lemma}[Uniform coupling]\label{lem:coupling}
Let $\mu$ be a distribution on $\{\pm1\}^m$ with $D_\infty(\mu) \le D$. There
is a coupling $(X,U)$ of $\mu$ and $\cU_m$ with the following properties:
\begin{enumerate}[label=(\alph*)]
  \item\label{it:coupling-distance}
    $\E\bigl[\dist(X,U)\bigr] \le \sqrt{mD}$.
    \item\label{it:coupling-tail}
    For every $I \subseteq [m]$ and $s > 0$, the following holds. For
    $u \in \{\pm1\}^I$, let $\nu_u$ be the distribution of $(X_{\bar I} \mid U_I = u)$.
    Then
    \begin{equation}\label{eq:tail}
      \Prb_{(X,U)}\Bigl[\nu_{U_I}(X_{\bar I}) > 2^{-(m-|I|)+D+s}\Bigr]
      \le 2^{-s}.
    \end{equation}
\end{enumerate}
\end{lemma}

\begin{proof}
For each $i \in [m]$, let
\[
  a_i(x_{<i}) \coloneqq \E_{X \sim \mu}\bigl[X_i \bigm| X_{<i} = x_{<i}\bigr].
\]
Take independent
$T_1, \ldots, T_m$, each uniform on $[-1,1]$, and for $i = 1, \ldots, m$ in turn,
set
\[
  X_i = 1 \iff T_i \le a_i(X_{<i}),
  \qquad
  U_i = 1 \iff T_i \le 0.
\]
Since $\Prb[T_i \le a] = (1+a)/2$, we have $X \sim \mu$ and $U \sim \cU_m$.
Moreover,
\[
  \Prb[X_i \ne U_i \mid X_{<i}] = \frac{|a_i(X_{<i})|}{2}.
\]

\paragraph{Part (a).}
Conditioned on $X_{<i} = x_{<i}$, the bit $X_i$ has distribution
$\operatorname{Ber}\bigl(\frac{1+a_i(x_{<i})}{2}\bigr)$. By the chain rule for
entropy and the elementary bound
$H\bigl(\operatorname{Ber}(\frac{1+a}{2})\bigr) \le 1 - a^2/4$ for
$a \in [-1,1]$,
\[
  H(X)
  = \sum_{i=1}^m H(X_i \mid X_{<i})
  \le m - \frac14 \sum_{i=1}^m \E\bigl[a_i(X_{<i})^2\bigr].
\]
Rearranging, and using $H(X) \ge H_\infty(X) = m - D_\infty(\mu) \ge m - D$,
we obtain
\begin{equation}\label{eq:coupling-bias-energy}
  \frac14 \sum_{i=1}^m \E\bigl[a_i(X_{<i})^2\bigr]
  \le m - H(X)
  \le D.
\end{equation}
Hence, by Cauchy--Schwarz and \eqref{eq:coupling-bias-energy},
\[
  \E\bigl[\dist(X,U)\bigr]
  = \frac12 \sum_{i=1}^m \E\bigl|a_i(X_{<i})\bigr|
  \le \sqrt{mD}.
\]

\paragraph{Part (b).}
Fix $I \subseteq [m]$ and let $Z \coloneqq (U_I, X_{\bar I})$. Since $U_I$ is
uniform,  
\[
  \E\bigl[2^{m-|I|}\,\nu_{U_I}(X_{\bar I})\bigr]
  = 2^{m-|I|} \sum_{u,y} \Prb[Z = (u,y)]\,
    \frac{\Prb[Z = (u,y)]}{\Prb[U_I = u]}
  = 2^m \sum_z \Prb[Z = z]^2.
\]
As $2^m \norm{\mu}_\infty \le 2^D$, Markov's inequality reduces
\eqref{eq:tail} to showing that the collision probability of $Z$ satisfies
\begin{equation}\label{eq:collision}
  \operatorname{col}(Z) \coloneqq \sum_z \Prb[Z = z]^2 \le \norm{\mu}_\infty.
\end{equation}

We prove \eqref{eq:collision} by induction on $m$. For $a\in \operatorname{supp}(X_1)$, let
$p_a \coloneqq \Prb[X_1 = a]$ and let $\mu_a$ be the distribution of
$(X_{>1} \mid X_1 = a)$. Then $\norm{\mu_a}_\infty \le \norm{\mu}_\infty/p_a$.

Note that, conditioned on any $(X_1, U_1) = (a,b)$ of positive probability,
the pair $(X_{>1}, U_{>1})$ is the coupling above for $\mu_a$. So by
induction, the conditional distribution of $Z_{>1}$ has collision probability at most
$\norm{\mu}_\infty/p_a$.

In the case $1 \notin I$, we have $Z_1 = X_1$, so
\[
  \operatorname{col}(Z)
  = \sum_{a} p_a^2 \operatorname{col}(Z_{>1} \mid X_1 = a)
  \le \sum_{a} p_a^2 \cdot \frac{\norm{\mu}_\infty}{p_a}
  = \norm{\mu}_\infty.
\]

If $1 \in I$, then $Z_1 = U_1$ is uniform. By convexity of collision
probability and the induction hypothesis,
\[
\begin{aligned}
  \operatorname{col}(Z)
  &= \frac12 \E_{U_1}\bigl[\operatorname{col}(Z_{>1} \mid U_1)\bigr]
   \le \frac12 \E_{X_1,U_1}\bigl[\operatorname{col}(Z_{>1} \mid X_1, U_1)\bigr] \\
  &\le \frac{\norm{\mu}_\infty}{2} \E_{X_1}\Bigl[\frac{1}{p_{X_1}}\Bigr]
   \le \norm{\mu}_\infty,
\end{aligned}
\]
where the last inequality follows from
$\E[1/p_{X_1}] = |\operatorname{supp}(X_1)| \le 2$. This completes the proof
of \eqref{eq:collision}, and hence of part~(b).
\end{proof}

\section{Proof of \Cref{thm:main}}
Throughout this section, $d \coloneqq \RR_{1/100}(F)$, and
$\rho \coloneqq c_0/\sqrt{d}$, where $c_0 > 0$ is an absolute constant small
enough that the bound in \Cref{thm:gstw} is at most $1/100$. For a
distribution $\mu$ on $\{\pm1\}^n$ and a set $J \subseteq [n]$, we write
$\mu_J$ for the marginal of $\mu$ on $J$.

\begin{lemma}[Planting a mostly agreeing block]\label{lem:planting}
Let $\mu \times \nu$ be a product distribution on $\{\pm 1\}^n \times \{\pm 1\}^n$ with
$\E_{\mu \times \nu} F \ge 1/2$. Let $J \subseteq [n]$ be a set of
$m \ge 100/\rho$ coordinates such that $\mu_J$ and $\nu_J$ have deficit at
most $D$, and set $b \coloneqq 20\rho\sqrt{mD}$.

Then there exist a set $I \subseteq J$ of size at least $\rho m/2$, a string
$w \in \{\pm1\}^I$, and a product distribution
$\widetilde\mu \times \widetilde\nu$, supported within the support of
$\mu \times \nu$, such that:
\begin{enumerate}[label=(\roman*)]
  \item every $(x,y) \in \supp(\widetilde\mu\times\widetilde\nu)$
    satisfies $\dist(x_I, w) \le b$ and $\dist(y_I, w) \le b$;
  \item $\E_{\widetilde\mu \times \widetilde\nu} F > 1/4$;    
  \item $\widetilde\mu_{J \setminus I}$ and $\widetilde\nu_{J \setminus I}$
    have deficit at most $D + 7$.
\end{enumerate}
\end{lemma}

\begin{proof}
Apply \Cref{lem:coupling} to $\mu_J$ and $\nu_J$ to obtain couplings
$(X_J, U_J)$ and $(Y_J, V_J)$, where $U_J$ and $V_J$ are uniform on $\{\pm1\}^J$.
Consider the following protocol $\Pi$ of cost $d$, in which Alice and Bob receive
$u, v \in \{\pm1\}^J$, respectively:
\begin{itemize}
  \item Alice samples $X_J$ from the coupling conditioned on $U_J = u$, and
    then samples $X_{[n]\setminus J}$ from $\mu$ conditioned on $X_J$.
  \item Bob does the same with $V_J = v$ to obtain $Y$, using independent
    randomness.
  \item They run the protocol for $F$ of cost $d$ and error at most $1/100$
    on $(X,Y)$.
\end{itemize}

We run $\Pi$ on inputs $(u,v)$ drawn from two different distributions. First,
let $(U_J', V_J')$ be independent and uniform, and let $X'$ and $Y'$ be the
strings that Alice and Bob sample on input $(U_J', V_J')$; then
$(X', Y') \sim \mu \times \nu$. Second, let $(U_J, V_J)$ be
$\rho$-correlated, and let $X$ and $Y$ be the strings sampled on input
$(U_J, V_J)$. Note that $X$ and $Y$ are not independent in this case.

By \Cref{thm:gstw}, the acceptance probabilities of $\Pi$ in the two runs
differ by at most $1/100$. Since the original protocol for $F$ errs with probability
at most $1/100$ on every input, we obtain
\begin{equation}\label{eq:planted-expectation}
  \E F(X,Y)
  \ge \E \Pi(U_J,V_J) - \frac{1}{100}
  \ge \E \Pi(U'_J,V'_J) - \frac{2}{100}
  \ge \E F(X',Y') - \frac{3}{100}
  \ge \frac{47}{100}.
\end{equation}

A $\rho$-correlated pair $(U_J,V_J)$ can be sampled by first choosing a random set
$I \subseteq J$ that includes each coordinate independently with probability
$\rho$, then sampling $V_I = U_I$ uniformly, and finally sampling
$U_{J\setminus I}$ and $V_{J\setminus I}$ independently and uniformly.

Write $R \coloneqq (I, U_I)$, let $p_R$ and $q_R$ be the conditional distributions of
$X_{J\setminus I}$ and $Y_{J\setminus I}$ given $R$, and put
$M_I \coloneqq 2^{-(m-|I|)+D+5}$. Consider the events
\[
  \begin{aligned}
  \mathcal A &\coloneqq
  \bigl\{\dist(X_I, U_I) \le b,\ p_R(X_{J\setminus I}) \le M_I\bigr\},\\
  \mathcal B &\coloneqq
  \bigl\{\dist(Y_I, V_I) \le b,\ q_R(Y_{J\setminus I}) \le M_I\bigr\}.
  \end{aligned}
\]
We call a value $(I_0, w)$ of $R$ \emph{good} if
\begin{equation}
\label{eq:goodness}
  |I_0| \ge \rho m/2
  \qquad\text{and}\qquad
  \E\bigl[F(X,Y)\,\mathbf{1}_{\mathcal A \cap \mathcal B} \bigm| R = (I_0, w)\bigr]
  > \frac14.
\end{equation}
We first show that a good value exists, and then use it to construct
$\widetilde\mu$ and $\widetilde\nu$.

\paragraph{A good $R$ exists.}
We bound the probability that $\mathcal A$ or $\mathcal B$ fails. Since $(X_J,U_J)$
is independent of $I$ and each
coordinate belongs to $I$ with probability $\rho$, by \Cref{lem:coupling}\,\ref{it:coupling-distance},
\[
  \E\bigl[\dist(X_I, U_I)\bigr]
  = \rho\,\E\bigl[\dist(X_J, U_J)\bigr]
  \le \rho\sqrt{mD},
\]
so, since $b = 20\rho\sqrt{mD}$, Markov's inequality gives
$\Prb[\dist(X_I, U_I) > b] \le 1/20$ (where we may assume $D > 0$).
Applying
\Cref{lem:coupling}\,\ref{it:coupling-tail} with $s = 5$ for each fixed value
of $I$ gives $\Prb[p_R(X_{J\setminus I}) > M_I] \le 1/32$. By the union bound, and the same argument applied to $(Y_J, V_J)$, we obtain 
\[
  \Prb[\mathcal A^c] \le \frac{1}{20} + \frac{1}{32} \quad \text{ and }\quad   \Prb[\mathcal B^c] \le \frac{1}{20} + \frac{1}{32}.
\]
Finally, by the Chernoff bound,
$\Prb[|I| < \rho m/2] \le e^{-\rho m/8} < 1/100$.

\smallskip
Let $\mathcal G \coloneqq \{|I| \ge \rho m/2\} \cap \mathcal A \cap \mathcal B$.
By the union bound, $\Prb[\mathcal G^c] < \frac15$, so by
\eqref{eq:planted-expectation},
\[
  \E\bigl[F(X,Y)\,\mathbf{1}_{\mathcal G}\bigr]
  \ge \frac{47}{100} - \frac15
  > \frac14.
\]
Averaging over $R$, some value $(I_0, w)$ of $R$ satisfies
$\E\bigl[F(X,Y)\,\mathbf{1}_{\mathcal G} \bigm| R = (I_0, w)\bigr] > \frac14$, and such a value satisfies the goodness condition \eqref{eq:goodness}.

\paragraph{Completing the proof.}
Fix a good value $(I_0, w)$ of $R$. Conditioned on $R = (I_0, w)$, the
inputs $X$ and $Y$ are independent. Let
\[
  \alpha \coloneqq \Prb\bigl[\mathcal A \bigm| R = (I_0, w)\bigr],
  \qquad
  \beta \coloneqq \Prb\bigl[\mathcal B \bigm| R = (I_0, w)\bigr],
\]
and let $\widetilde\mu$ and $\widetilde\nu$ be the distributions of $X$ given
$R = (I_0, w)$ and $\mathcal A$, and of $Y$ given $R = (I_0, w)$ and
$\mathcal B$, respectively.

Property (i) holds by the definition of $\mathcal A$ and $\mathcal B$, since
$U_{I_0} = V_{I_0} = w$. For (ii), note that
\[
  \frac14
  < \E\bigl[F(X,Y)\,\mathbf{1}_{\mathcal A \cap \mathcal B} \bigm| R = (I_0, w)\bigr]
  = \alpha\beta\, \E_{\widetilde\mu\times\widetilde\nu} F
  \le \min\Bigl\{\alpha,\ \beta,\ \E_{\widetilde\mu\times\widetilde\nu} F\Bigr\},
\]
which also gives $\alpha, \beta > 1/4$.

For (iii), let $p$ be the distribution of $X_{J\setminus I_0}$ given
$R = (I_0, w)$. Then every $z \in \{\pm1\}^{J\setminus I_0}$ satisfies
\[
  \widetilde\mu_{J\setminus I_0}(z)
  \le \frac{p(z)\,\mathbf 1_{\{p(z) \le M_{I_0}\}}}{\alpha}
  < 4M_{I_0}
  = 2^{-(m-|I_0|)+D+7},
\]
and the same holds for $\widetilde\nu$.
\end{proof}

\begin{lemma}[Restoring the expectation]\label{lem:restoring-expectation}
Let $\xi$ be any distribution with $\E_\xi F \geq 1/4$. Then there is a rectangle
$A \times B$ such that
\[
    \xi(A \times B) \ge \frac{2^{-d}}{5}
    \qquad\text{and}\qquad
    \E_\xi [ F \mid A \times B ] \ge \frac{1}{2} .
\]
\end{lemma}
\begin{proof}
Fix the randomness of a protocol for $F$ of cost $d$ and error at most $1/100$ to obtain a
deterministic protocol with probability of error at most $1/100$ under $\xi$.  Let $\mathcal L$
be the set of leaves on which it outputs $1$; recall that each leaf is a rectangle $A \times B$.
Without loss of generality, we may assume that $\E_\xi [F \mid L] \geq 1/2$ for all $L \in
\mathcal L$ (since otherwise we may flip the output of the protocol on $L$ while reducing the
probability of error). Since the probability of error is at most $1/100$ and $\E_\xi F \geq 1/4$,
we have $\sum_{L \in \mathcal L} \xi(L) \geq \frac{1}{4} - \frac{1}{100} \geq \frac{1}{5}$.
Since $|\mathcal L| \leq 2^d$, there is a leaf $L$ with $\xi(L) \geq 2^{-d}/5$.
\end{proof}

\begin{remark}\label{rem:deficit} Applying \cref{lem:restoring-expectation}
with a product distribution $\mu \times \nu$ and
conditioning on $A \times B$ gives the product distribution
$\mu(\,\cdot \mid A) \times \nu(\,\cdot \mid B)$. Since
$\mu(A), \nu(B) \ge 2^{-d}/5$, conditioning multiplies every probability by
at most $5 \cdot 2^d$. Hence, for every $J \subseteq [n]$, the deficits of
the marginals on $J$ increase by at most $d + \log 5 < d + 3$.
\end{remark}

\begin{proof}[Proof of \Cref{thm:main}]
As in the overview, assume $\E_{\cU_n\times\cU_n}F\ge1/2$. Note that $d\ge1$,
and suppose, towards a contradiction, that $d\le c_\gamma n^{2/3}$ for a
sufficiently small constant $c_\gamma>0$.

For each step $t \in \{0, \dotsc, T\}$ (with $T$ determined below), we maintain a product
distribution $\mu_t\times\nu_t$ with $\E_{\mu_t\times\nu_t}F\ge1/2$, a set $K_t\subseteq[n]$ of
planted coordinates with a reference string $a_t\in\{\pm1\}^{K_t}$, and numbers $\Delta_t$ and $D_t$
such that every $(x,y)$ in the support satisfies $\dist(x_{K_t},a_t),\dist(y_{K_t},a_t)\le\Delta_t$,
and the marginals of $\mu_t$ and $\nu_t$ on $J_t\coloneqq[n]\setminus K_t$ have deficit at most
$D_t$. Write $m_t\coloneqq|J_t|$. Initially, $K_0=\varnothing$, $\mu_0=\nu_0=\cU_n$, and
$\Delta_0=D_0=0$.

As long as $m_t>\gamma n/2$, we apply \Cref{lem:planting} with $J=J_t$, followed by
\Cref{lem:restoring-expectation}, and add the new block $I$ to $K_t$ to obtain $K_{t+1}$, and we
extend $a_t$ by $w$ to obtain $a_{t+1}$. Since both lemmas only shrink the support, the distance
bounds on $K_t$ are preserved, and by \Cref{lem:planting}~(iii) and \Cref{rem:deficit},
\[
 m_{t+1}\le(1-\rho/2)\,m_t,
 \qquad
 D_{t+1}\le D_t+d+10,
 \qquad
 \Delta_{t+1}\le\Delta_t+20\rho\sqrt{m_tD_t}.
\]
Consequently, the process stops after
$T\le\lceil(2/\rho)\ln(2/\gamma)\rceil=O_\gamma(1/\rho)$ rounds. Moreover,
$D_t\le t(d+10)$ and $m_t\le n$, so
\[
 \Delta_T
 \le20\rho\sqrt{n(d+10)}\,T^{3/2}\\
 =O_\gamma\Bigl(\sqrt{n(d+10)/\rho}\Bigr)
 =O_\gamma\bigl(\sqrt n\,d^{3/4}\bigr).
\]
Since $\rho m_t>c_0\gamma n/(2\sqrt d)\ge c_0\gamma n^{2/3}/(2\sqrt{c_\gamma})$
before stopping, the condition $\rho m_t\ge100$ of \Cref{lem:planting} holds
for sufficiently large $n$.

By choosing $c_\gamma$ sufficiently small, we obtain $\Delta_T\le\gamma n/4$.
Thus every pair $(x,y)$ in the final support satisfies
\[
 \dist(x,y)\le 2\Delta_T+m_T\le\gamma n,
\]
forcing $F(x,y)=0$. This contradicts $\E_{\mu_T\times\nu_T}F\ge1/2$. Hence
$\RR_{1/100}(F)=\Omega_\gamma(n^{2/3})$.
\end{proof}

\bigskip

\subsection*{Acknowledgements}
H.H. wishes to thank Amir Yehudayoff and Arkadev Chattopadhyay for earlier discussions on attempts to solve this problem.

M.G. and W.Y. are supported by the Swiss State
Secretariat for Education, Research and Innovation
(SERI) under contract number MB22.00026.
H.H. and N.H. are supported by NSERC Discovery Grants.
Y.D. and S.L. are supported by Simons Investigator Award \#929894. S.L. is also supported by NSF Award CCF-2425349.

\bigskip

\DeclareUrlCommand{\Doi}{\urlstyle{sf}}
\renewcommand{\path}[1]{\small\Doi{#1}}
\renewcommand{\url}[1]{\href{#1}{\small\Doi{#1}}}
\bibliographystyle{alphaurl}
\bibliography{refs} 

\bigskip 

\authoraddress
  {Yogesh Dahiya}
  {University of California San Diego,
   La Jolla, California, USA}
  {ydahiya@ucsd.edu}

\authoraddress
  {Mika G\"o\"os}
  {EPFL, Lausanne, Switzerland}
  {mika.goos@epfl.ch}

\authoraddress
  {Nathaniel Harms}
  {University of British Columbia,
   Vancouver, Canada}
  {nharms@cs.ubc.ca}

\authoraddress
  {Hamed Hatami}
  {McGill University, Montr\'eal, Canada}
  {hatami@cs.mcgill.ca}

\authoraddress
  {Pooya Hatami}
  {The Ohio State University,
   Columbus, Ohio, USA}
  {hatami.2@osu.edu}

\authoraddress
  {Shachar Lovett}
  {University of California San Diego,
   La Jolla, California, USA}
  {slovett@ucsd.edu}

\authoraddress
  {Weiqiang Yuan}
  {EPFL, Lausanne, Switzerland,
   and University of Copenhagen,
   Copenhagen, Denmark}
  {weiqiang.yuan@di.ku.dk}

\end{document}